\documentclass[a4paper,12pt]{article}
\usepackage[utf8]{inputenc}
\usepackage{amssymb, mathtools}
\usepackage{algpseudocode,algorithm}
\usepackage[margin=2.00cm]{geometry}
\usepackage{tabularx}
\usepackage[numbers]{natbib}
\usepackage{setspace}
\usepackage{fancybox}
\usepackage{tikz}
\usepackage{rotating}
\usepackage{censor}

\newcommand{\wrt}[1]{\mathrm{d}{#1}}
\newcommand{\bs}[1]{\boldsymbol{#1}}

\title{Multi-level Approximate Bayesian Computation}
\author{Christopher Lester\footnote{Mathematical Institute, Woodstock Road, Oxford, OX2 6GG, UK. Email: \texttt{lesterc AT maths.ox.ac.uk.}}}
\date{7 August 2019}

\begin{document}

\maketitle

\begin{abstract}Approximate Bayesian Computation (ABC) is widely used to infer the parameters of discrete-state continuous-time Markov networks. In this work, we focus on models that are governed by the Chemical Master Equation (the CME). Whilst originally designed to model biochemical reactions, CME-based models are now frequently used to describe a wide range of biological phenomena mathematically. We describe and implement an efficient \emph{multi-level} ABC method for investigating model parameters. In short, we generate sample paths of CME-based models with varying time resolutions. We start by generating low-resolution sample paths, which require only limited computational resources to construct. Those sample paths that compare well with experimental data are selected, and the temporal resolutions of the chosen sample paths are recursively increased. Those sample paths unlikely to aid in parameter inference are discarded at an early stage, leading to an optimal use of computational resources. The efficacy of the \emph{multi-level} ABC is demonstrated through two case studies.

\end{abstract}

\section{Introduction} \label{__label__2976342c969e4c8e8af17a17f45cf902}
Well-designed mechanistic models can provide insights into biological networks that are imperceptible to machine learning techniques. For example, where suitable experimental data exists, mechanistic models can often provide strong evidence for the causative relationships that underpin a given biological model~\citep{__ref__56fdd8bc574d4653a485c3e0b022ed73}. Mechanistic, multi-scale models have been able to assimilate physical features and behaviours that span multiple time- and spatial-scales~\citep{__ref__37125c81985a4d0c850af6cf087b802b}. 

To draw reliable conclusions from a mechanistic model, experimental data are often used to inform the planning, implementation and usage of the model. In particular, inference methods are frequently used to choose numerical parameters for a given model, or perhaps to select the most appropriate description from a class of possible models. It thus essential that uncertainties in inferred parameters, and hence, in any conclusions, are accurately quantified. 

In this work, we focus on discrete state-space stochastic models that are governed by the Chemical Master Equation (the CME). Whilst originally designed to model biochemical reactions, CME-based models are now frequently used to describe a wide range of biological phenomena mathematically~\citep{__ref__6f6fbb37efec440e98cb3ec3d5482cd6, __ref__2c37ad24e6df4cc994c138ecafd7b796, __ref__f1a604725ef54db7a3d7727502e017a0}. For CME-based models, Approximate Bayesian Computation (ABC) is becoming an increasingly popular method of inferring model parameters~\citep{__ref__3b73781bb5284517aa5407b8c3de9021, __ref__3d065346aeef453e887da030e55cb70e, __ref__d2e49582513e47d99546269ad3c1a365}.

Bayesian methods are widely used for parameter inference. In this setting, any existing knowledge of a parameter, $\theta$, is encoded as a \emph{prior} distribution, $\pi(\theta)$; the probability of a parameter, $\theta$, given data $D$, $\mathbb{P}[\theta \mid D]$ is then recorded in the \emph{posterior} distribution~\citep{__ref__080d917a28b643d988808b9d6746b5c7}. Following Bayes' theorem, the \emph{likelihood}, $\mathbb{P}[D \mid \theta]$, relates the \emph{posterior} distribution to the \emph{prior}: \begin{equation*}\mathbb{P}[\theta \mid D] \propto \mathbb{P}[D \mid \theta] \times \pi(\theta).\end{equation*}

For many mechanistic models, the likelihood is intractable, rendering many Bayesian inference methods infeasible~\citep{__ref__d2e49582513e47d99546269ad3c1a365}. As a likelihood-free method, ABC is unaffected and is often able to infer parameters with a high degree of accuracy. ABC uses stochastic simulation to infer parameters: after sampling parameters from a prior distribution, sample paths or realisations of the model of interest are generated. By studying the sample paths that have been generated, the posterior distribution is estimated. 

In this work, we describe and implement an efficient \emph{multi-level} ABC method for investigating model parameters. In short, we generate sample paths of CME-based models with varying time resolutions. We will first generate low-resolution sample paths, by which we mean sample paths with few time-steps (and therefore low-quality approximations of the model dynamics), and we will later generate high-resolution sample paths, by which we mean sample paths with more time-steps (but which require more computational resources to generate). The \emph{multi-level} ABC (`ML-ABC') method that we set out in this manuscript proceeds by:
\begin{enumerate}
 \item Starting by generating low-resolution sample paths to infer model parameters.
 \item Then, choosing a subset of sample paths, with a view to improving their time resolution. The subset is chosen to include more of the sample paths likely to influence the posterior distribution. The sample paths not included in the subset are discarded.
 \item Step 2 is recursively repeated with the remaining sample paths, until all sample paths have a sufficiently high resolution. The posterior distribution is then estimated. 
\end{enumerate} The result is that we can discard many (if not most) sample paths after a quickly-generated low-resolution version is generated. The bulk of the computational effort is thus expended on improving the time resolution of those sample paths that most likely to contribute to the posterior distribution.

This work is predicated on, and seeks to unify, three principal previous works: \begin{enumerate}
\item Whilst rejection sampling had previously been used to sample a posterior distribution~\citep{__ref__44faf37082bb4134b3a03a1a36435826}, \citet{__ref__d2e49582513e47d99546269ad3c1a365} provide perhaps the most widely-read account of the ABC inference framework, and we build on their framework.
\item \citet{__ref__fda9da77c68d4bb891679909cb57ed58} describes a `lazy' ABC approach. In this work, we apply the theory developed by \citet{__ref__fda9da77c68d4bb891679909cb57ed58} in a novel way.
\item \citet{__ref__db063527d9b94dcca12e9397daa3c169} describes a multi-level method for calculating Monte Carlo estimates. A recent work~\citep{__ref__a42fe24004244e1ca2f0aad5851f3653} discusses the means to re-use random inputs to generate sample paths with different time resolutions. We follow this approach.
\end{enumerate}

This manuscript is arranged as follows: in Section \ref{__label__6b90150fcf9547afa4c1f2a5cac7059f} we provide background material covering biochemical reaction networks, and the ABC algorithm. In Section \ref{__label__659100d1f2c14b66a0f4493f0729ea9b} our new framework is set out. The framework is implemented and tested in Section \ref{__label__1d1140eeb1534e8b956d79bc5f640203}, and a discussion of our results then follows. 

\section{Inference for stochastic biochemical networks} \label{__label__6b90150fcf9547afa4c1f2a5cac7059f}
This section is in two parts: first, we explain the CME modelling framework in Section \ref{__label__bca8b43c50a94ed688a3a952869d090f}, and in Section \ref{__label__6aee109ee913466e867f66a6f73065f4}, the Approximate Bayesian Computation methodology is outlined. 

\subsection{Biochemical reaction networks} \label{__label__bca8b43c50a94ed688a3a952869d090f}
We study a biochemical network comprising $N$ species, $S_1, \dots, S_N$, that may interact through $M$ reaction channels, $R_1$,$\dots$,$R_M$. In our case, the dynamics of a biochemical reaction network are governed by the Chemical Master Equation (CME)~\citep{__ref__d45fb4c000644fc9bf5bd194cba3ab20}. At time $t$, the population, or copy number, of species $S_i$ is denoted by $X_i(t)$, and the state vector, $\bs{X}(t)$, is given by \begin{equation}
\bs{X}(t) \coloneqq \left[X_1(t), \dots, X_N(t)\right]^T. \label{__label__9a73ab33264847b5bd60742aaf26dc86} \end{equation}

We associate two quantities with each reaction channel $R_j$. The first is the stoichiometric or state-change vector, \begin{equation}
\bs{\nu}_j \coloneqq \left[\nu_{1j}, \dots, \nu_{Nj}\right]^T,  \label{__label__e683816cc0f047f88a272bba19821b7f}
\end{equation}
where $\nu_{ij}$ is the change in the copy number of $S_i$ caused by reaction $R_j$ taking place. The second quantity is the propensity function, $p_j(\bs{X}(t))$. For infinitesimally small $\wrt{t}$, the rate $p_j(\bs{X}(t))$ is defined as follows: \begin{equation*}
p_j(\bs{X}(t))\wrt{t} \coloneqq \mathbb{P}\left[R_j \text{ occurs in } [t, t + \wrt{t}) \right].
\end{equation*} We assume that the system is well-stirred, so the reaction activity can be modelled with mass action kinetics. In this case, the propensity function of reaction $R_j$, $p_j$, is proportional to the number of possible combinations of reactant particles in the system~\citep{__ref__d45fb4c000644fc9bf5bd194cba3ab20}. The constants of proportionality are known as rate constants and must be \emph{inferred} from experimental data~\citep{__ref__f1a604725ef54db7a3d7727502e017a0}.

For our purposes, it is convenient to use the Random Time Change Representation (RTCR) of the CME framework, which was first described by \citet{__ref__a7750d6ef7e5470197d652a81740b871}. The RTCR describes the dynamics of a biochemical network by using a set of unit-rate Poisson processes. The number of times reaction $R_j$ (for $j = 1, \dots, M$) takes place (`fires') over the time interval $(0, T]$ is given by a Poisson counting process \begin{equation*}
 \mathcal{Y}_j \left(0, \int_0^T p_j(\bs{X}(t)) \wrt{t}\right),
\end{equation*} where $\mathcal{Y}_j$ is a unit-rate Poisson process, and $\mathcal{Y}_j(\alpha, \beta)$ is defined as \begin{equation}
\mathcal{Y}_j(\alpha, \beta) \coloneqq \text{\# of arrivals in } (\alpha, \beta].
\end{equation} Every time reaction $R_j$ occurs, the state vector (see Equation \eqref{__label__9a73ab33264847b5bd60742aaf26dc86}), is updated by adding the appropriate stoichiometric vector (see Equation \eqref{__label__e683816cc0f047f88a272bba19821b7f}) to it. Therefore, by considering all possible reactions over the time interval $(0,T]$, we can determine the state vector at time $T$ as \begin{equation}
 \bs{X}(T) = \bs{X}(0) + \sum_{j=1}^{M}  \mathcal{Y}_j \left(0,\int_0^T p_j(\bs{X}(t)) \wrt{t}\right) \cdot \nu_j. \label{__label__84d2b77cf3bf4118b0e534292f4a7a43}
\end{equation}  
We can think of different sets of Poisson processes, $\{\mathcal{Y}_j$ \text{ for }$j = 1, \dots, M\}$, as providing the randomness for different sample paths of our biochemical reaction network. Each random sample path is thus uniquely associated with a set of Poisson processes (and each reaction channel is associated with a specific Poisson process). 

It is possible to directly simulate the sample path described by Equation \eqref{__label__84d2b77cf3bf4118b0e534292f4a7a43} by using the Modified Next Reaction Method (the MNRM)~\citep{__ref__8f714e91c9204244b49792e2f2ea4fa8}. The MNRM is set out as Algorithm \ref{__label__ee5b7b6d2e204309b2a457c609814a7a}; the sample paths produced by this method are statistically indistinguishable from the sample paths generated with the famed Gillespie Algorithm. As with the Gillespie Algorithm, the MNRM is a serial algorithm: it uses a relatively high level of computational resources to generate each sample path as it `fires' only a single reaction at each step of the algorithm. As alluded to above, a feature of the MNRM is that it associates a specific Poisson process with each reaction channel: we will rely on this feature throughout the remainder of this manuscript.

\begin{algorithm}[bth]
\caption{\protect\vphantom{$A^A$}The MNRM. A single sample path is generated by using Poisson processes, $\mathcal{Y}_j$ (for $j=1, \dots, M$), to `fire' reactions through different reaction channels.\protect\vphantom{$A_A$}}
\label{__label__17a38d1ae71849959c21f8e81c665c2b}
\begin{algorithmic}[1]
  \Require initial conditions, $\bs{X}(0)$, and terminal time, $T$.\protect\vphantom{$A_A^A$} 
  \State set $\bs{X} \leftarrow \bs{X}(0)$, and set $t \leftarrow 0$ 
  \State for each $R_j$, set $P_j \leftarrow 0$, generate $T_j \leftarrow \text{Exp}(1)$ \Comment{\textit{$T_j$ is the first inter-arrival time of $\mathcal{Y}_j$.}} 
  \Loop
  \State for each $R_j$, calculate propensity values $p_j(\bs{X})$ and calculate $\Delta_j$ as \begin{equation*} \Delta_j = \frac{T_j - P_j}{p_j} \end{equation*}
  \State set $\Delta \leftarrow \min_j \Delta_j$, and $k \leftarrow \text{argmin}_j \Delta_j$
  \If{$t + \Delta > T$}
  \State \textbf{break}
  \EndIf
  \State set $\bs{X}(t + \Delta) \leftarrow \bs{X}(t) + \bs{\nu}_k$, set $t \leftarrow t + \Delta$, and for each $R_j$, set $P_j \leftarrow P_j + p_j \cdot \Delta$
  \State generate $u \sim \text{Exp}(1)$, then set $T_k \leftarrow T_k + u$ \Comment{\textit{$u$ is the next inter-arrival time of $\mathcal{Y}_k$.}} 
  \EndLoop
\end{algorithmic} \label{__label__ee5b7b6d2e204309b2a457c609814a7a}
\end{algorithm} 

Sample paths of $\bs{X}(t)$ can be generated more efficiently by using the tau-leap assumption. In this case, the state vector, $\bs{X}(t)$, is only updated at fixed times: for example, at times $t = \tau, 2\cdot\tau, \dots$, where $\tau$ is an appropriate time-step. The tau-leap assumption means that the following substitution can be made in Equation \eqref{__label__84d2b77cf3bf4118b0e534292f4a7a43}:
\begin{equation}
 \int_0^T p_j(\bs{X}(t)) \wrt{t} \rightarrow \sum_{k=0}^{K} p_j(\bs{X}(k\cdot\tau))\cdot \tau. 
\end{equation} Equation \eqref{__label__84d2b77cf3bf4118b0e534292f4a7a43} can then be rearranged into an update formula: 
\begin{equation}
 \bs{X}(K \cdot \tau) = \bs{X}((K-1)\cdot \tau) + \sum_{j=1}^M \mathcal{Y}_j \left( \sum_{k=0}^{K-1} p_j(\bs{X}(k\cdot\tau))\cdot \tau, \sum_{k=0}^{K} p_j(\bs{X}(k\cdot\tau))\cdot \tau \right) \cdot \nu_j. \label{__label__3576a62258ac4e9d8953e5bbea204ecf}
\end{equation} Effectively, Equation \eqref{__label__3576a62258ac4e9d8953e5bbea204ecf} states that, to advance from time $t = (K-1)\cdot \tau$ to time $t = K \cdot \tau$, we `read-off' the number of times each Poisson process, $\mathcal{Y}_j$ (for $j = 1, \dots, M$), has `fired' during the time-interval $\left(\sum_{k=0}^{K-1} p_j(\bs{X}(k\cdot\tau))\cdot \tau, \sum_{k=0}^{K} p_j(\bs{X}(k\cdot\tau))\cdot \tau\right]$. As expected, the length of each interval is $p_j\left(\bs{X}((K-1)\cdot\tau)\right)\cdot\tau$. 

As explained above, each sample path is associated with a unique set of Poisson processes, $\{\mathcal{Y}_j$ \text{ for }$j = 1, \dots, M\}$. The same set of Poisson processes can be used, but with a different value of $\tau$, to change the resolution at which a sample path is constructed. 

\begin{figure}[t]

\centering
\includegraphics[width=.75\linewidth]{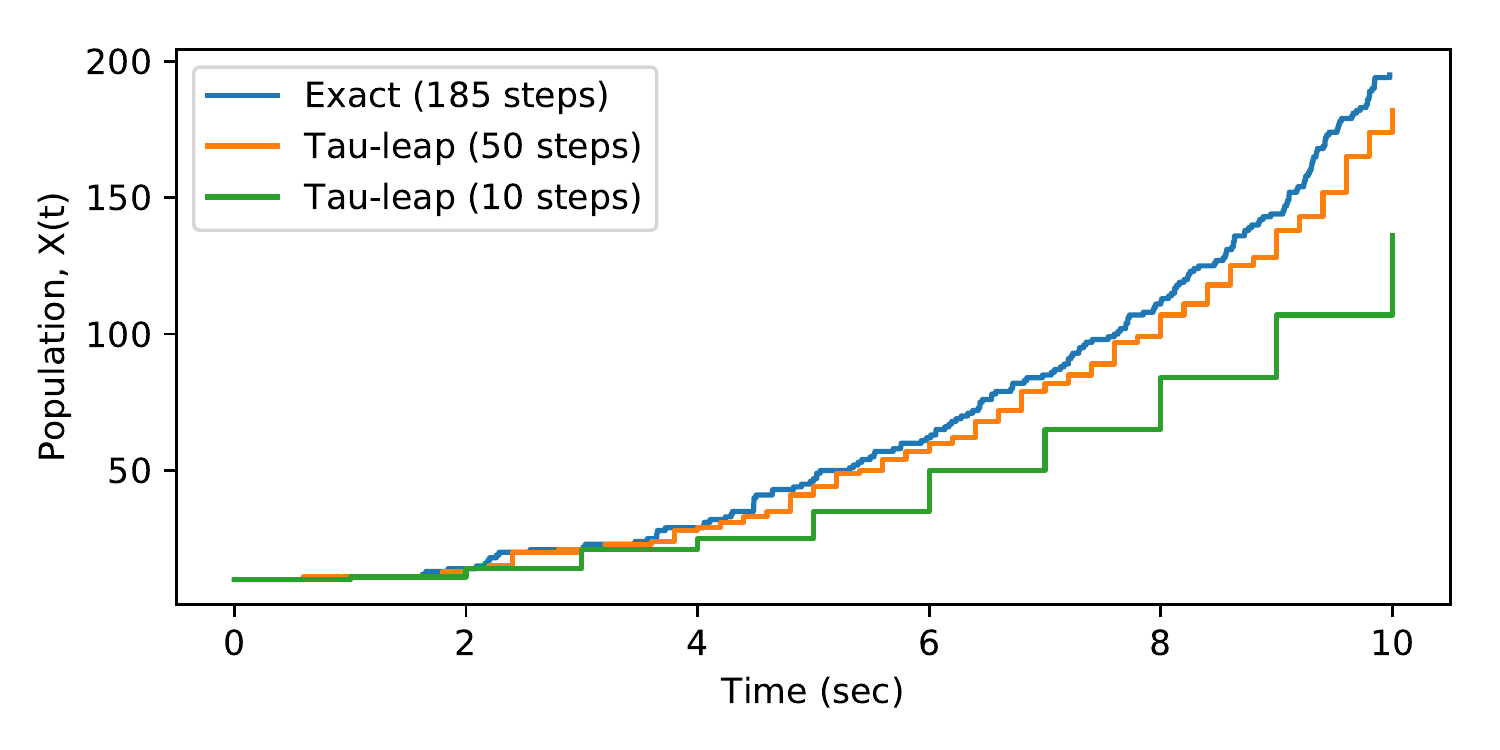} 
\caption{A single sample path of System \eqref{__label__5da48c70d96c4ee39993f234ef1294eb} is shown, but which has been generated by using different time-steps in Equation \eqref{__label__3576a62258ac4e9d8953e5bbea204ecf}. We show sample paths with $\tau = 1.0$ (10 steps) and $\tau = 0.2$ (50 steps), together with the MNRM sample path ($\tau \downarrow 0$). The same Poisson process is used throughout.} \label{__label__4ccf8701d0014669b51583e6f21de795}

\end{figure}

We illustrate with an example:

\textbf{Case study.} Consider a reaction network comprising a single species, $X$, together with a reaction channel, $R_1$, which we specify as: \begin{equation} R_1: X \xrightarrow{\theta} X + X. \label{__label__5da48c70d96c4ee39993f234ef1294eb} \end{equation}We take $X(0) = 10$ and $\theta = 0.3$. In Figure \ref{__label__4ccf8701d0014669b51583e6f21de795}, we use a single Poisson process, $\mathcal{Y}$, to generate a single sample path, but at different resolutions. As $\tau \downarrow 0$, the sample path corresponds with a sample path generated by the MNRM.

\subsection{Approximate Bayesian Computation} \label{__label__6aee109ee913466e867f66a6f73065f4}
We now detail how ABC infers the parameters of the CME-based stochastic model described in Section \ref{__label__bca8b43c50a94ed688a3a952869d090f}. In Section \ref{__label__2976342c969e4c8e8af17a17f45cf902} we explained that the ABC method is an example of a likelihood-free inference method that allows us to directly estimate the posterior, $\mathbb{P}[\theta \mid D]$: \begin{itemize}
\item The prior, $\pi(\cdot)$, encodes any existing knowledge regarding the parameters. An uninformative prior -- for example, a uniform distribution -- can be used if little is known about the model. 
\item A parameter\footnote{Whilst we refer to `parameter' as singular, indeed, the parameter could be a vector.}, $\theta$, is sampled from the prior. Then, based on this parameter choice, a sample path, $\bs{X}$, is generated. We write $\bs{X} \sim f(\cdot |\theta)$ to indicate this. This step can be repeated $N$ times to generate many sample paths. 
\item The sample paths, $\{\bs{X}_1, \dots, \bs{X}_N\}$, are compared with the observed data, which we label $\widehat{\bs{X}}$. As the CME model is stochastic, the observed data are very unlikely to coincide with a generated sample path.

Therefore, summary statistics are specified, and the summary statistics of sample paths are compared with the summary statistics of the raw data. For example, we might take $s(\bs{X}) = \left[\bs{X}(t_1), \dots, \bs{X}(t_K)\right]$, and then compare summary statistics with the Frobenius norm for matrices, $\|s(\bs{X}) - s(\widehat{\bs{X}})\|_2$. 

\item A tolerance, $\varepsilon > 0$, is specified. Those sample paths that match the data -- for example, where $\|s(\bs{X}) - s(\widehat{\bs{X}})\| < \varepsilon$ -- are \emph{accepted}, whilst the remainder are \emph{rejected}. The parameters associated with the accepted sample paths are used to empirically generate an approximate posterior distribution, $\mathbb{P}[\theta \mid \|s(\bs{X}) - s(\widehat{\bs{X}})\| < \varepsilon]$. Where the summary statistics are \emph{sufficient}, as $\varepsilon \downarrow 0$, the true posterior, $\mathbb{P}[\theta \mid \bs{X}]$, can be recovered~\citep{__ref__3b73781bb5284517aa5407b8c3de9021}.
\end{itemize}

This is the rejection ABC algorithm, and it is stated as Algorithm \ref{__label__678f924617094025815fa9832fc0f45c}. This basic algorithm can be improved through the use of a Markov-Chain Monte Carlo (MCMC) approach~\citep{__ref__7a131cea9f2649358beff55cb718226b}, or through a Sequential Monte Carlo sampler method~\citep{__ref__d2e49582513e47d99546269ad3c1a365}.

 \begin{algorithm}[bth]
\caption{Rejection ABC\protect\vphantom{$A_A^A$}. The loop is repeated to generate sample paths, $\bs{X} \sim f(\cdot \mid \cdot)$, until sufficiently many values of $\theta$ have been accepted. }
\label{__label__678f924617094025815fa9832fc0f45c}
 \begin{algorithmic}[1]
  \Require data, $\widehat{\bs{X}}$, tolerance, $\varepsilon$, summary statistic, $s(\cdot)$, distance metric $\|\cdot\|$, and prior, $\pi(\cdot)$\protect\vphantom{$A_A^A$}
\Loop
  \State sample $\theta \sim \pi(\cdot)$
  \State generate sample path $\bs{X} \sim f(\cdot \mid \theta)$
  \If{$\|s(\bs{X}) - s(\widehat{\bs{X}})\| < \varepsilon$}
  \State accept $\theta$
  \EndIf

  \EndLoop
 \end{algorithmic}
\end{algorithm}

Before proceeding to Section \ref{__label__659100d1f2c14b66a0f4493f0729ea9b}, and in following \citet{__ref__fda9da77c68d4bb891679909cb57ed58}, we point out that Algorithm \ref{__label__678f924617094025815fa9832fc0f45c} can be viewed as a weighted sampling algorithm. In particular, after each parameter, $\theta$, is sampled, it can be given a weight, $\omega$, according to \begin{equation*}
\omega = \begin{cases} 
          1 & \text{ if } \bs{X} \sim f(\cdot \mid \theta) \text{ is accepted.} \\
          0 & \text{ if } \bs{X} \sim f(\cdot \mid \theta) \text{ is rejected.} 
         \end{cases}
         \end{equation*}
This opens up at least two possibilities: firstly, one can sample $\theta$ as $\theta \sim g(\cdot)$, where $g(\cdot)$ has been carefully chosen to explore the parameter space most likely to result in a parameter being accepted. The weight of $\theta$ is then scaled by a factor of $\pi(\theta) / g(\theta)$~\citep{__ref__d2e49582513e47d99546269ad3c1a365}. The second possibility involves terminating the simulation of $\bs{X}$ prematurely: we explore this idea in Section \ref{__label__659100d1f2c14b66a0f4493f0729ea9b}. The use of non-integer weights means that an \emph{effective sample size} needs to be calculated. Kish's \emph{effective sample size}~\citep{__ref__b2a67b6368964e999efcd52bd4bf52eb} is given by \begin{equation}
\text{ESS} = \frac{\left[\sum_{r=1}^N \omega_r\right]^2}{\vphantom{A^{A^{A^A}}}\sum_{r=1}^N \omega_r^2}. \label{__label__045cbdb1be72461b96f5c0175bb68360}
\end{equation}

Having outlined the CME modelling and ABC inference frameworks, we are now in a position to set out our new \emph{multi-level} ABC (`ML-ABC') method.

\section{Multi-level Approximate Bayesian Computation} \label{__label__659100d1f2c14b66a0f4493f0729ea9b}
This section is in three parts: first, in Section \ref{__label__0ae2c4ede8e94b68a908d2e812d70329}, we discuss and adapt  \citet{__ref__fda9da77c68d4bb891679909cb57ed58}'s `lazy' ABC method; then, in Section \ref{__label__b27923208d3d4b08988f3e77aa5a3cb0} we refer to an earlier work \citep{__ref__a42fe24004244e1ca2f0aad5851f3653} to explain how to generate the requisite sample paths. Finally, Section \ref{__label__786cad91bb664bc78e7876eda9a9a892} spells out our method.

\subsection{Early-rejection ABC} \label{__label__0ae2c4ede8e94b68a908d2e812d70329}
\citet{__ref__fda9da77c68d4bb891679909cb57ed58} introduces what we will call an `early-rejection' ABC method, which is now restated. Let $\bs{X}$ refer to a sample path of the model of interest, and recall that $\bs{X}$ is sampled as $\bs{X} \sim f(\cdot \mid \theta)$. Let $\bs{Y}$ refer to a \emph{partial} sample path of the model that meets the following requirements: \begin{itemize}
\item $\bs{Y}$ can be sampled as $\bs{Y} \sim g(\cdot \mid \theta)$ cheaply. Let $\bs{\omega}$ be the random noise that generates $\bs{Y}$. 
\item Given the noise associated with the partial sample, $\bs{\omega}$, a full sample can be deduced as $\bs{X} \sim f(\cdot \mid \theta, \bs{\omega})$.
\end{itemize} The early-rejection procedure is as follows: a sample path $\bs{Y}$ is generated. We let a decision function $\alpha(\bs{Y})$ provide a \emph{continuation probability}: with probability $\alpha(\bs{Y})$, the partial sample $\bs{Y}$ is upgraded to a full sample, $\bs{X}$, by simulating $\bs{X} \sim f(\cdot \mid \theta, \bs{\omega})$. The weight of the full sample is then re-scaled by a factor of $1/\alpha( \bs{Y})$, and the ABC algorithm proceeds. Alternatively, if the partial sample, $\bs{Y}$, is not upgraded, it is discarded.

\citet{__ref__fda9da77c68d4bb891679909cb57ed58} envisages a range of scenarios where early-rejection ABC can be implemented. For example, suppose that the sample path $\bs{X}$ is generated over a time-interval $[0, t)$. Then the partial sample path, $\bs{Y}$, could be generated over a shorter time-interval $[0, s)$ (with $s \ll t$, or with $s$ a random stopping time). Based on the shorter sample path, the probability of the complete sample path meeting the ABC acceptance criterion, $\mathbb{P}\left[\|s(\bs{X}) - s(\widehat{\bs{X}}) \| < \varepsilon \mid \theta, \bs{\omega} \right]$, can be estimated. Unlikely sample paths can thus be discarded after using only a small amount of computational resources, but at the cost an increased variance.

Asymptotically, the optimal efficiency is achieved where the ratio of the expected effective sample size (see Equation \eqref{__label__045cbdb1be72461b96f5c0175bb68360}) to the expected CPU time per sample, $\mathbb{E}[T]$, is maximised: \begin{equation}
\dfrac{\mathbb{E}[\omega]^2}{\mathbb{E}[\omega^2] \cdot \mathbb{E}[T]}. \label{__label__c5426463d79b428197b24a63f7b08b9b}
\end{equation}

We aim to choose the continuation probability function, $\alpha(\bs{Y})$, so that Equation \eqref{__label__c5426463d79b428197b24a63f7b08b9b} is as large as possible. Diverging slightly from \citet{__ref__fda9da77c68d4bb891679909cb57ed58}, we let $\phi(\bs{Y})$ be a (vector) summary statistic\footnote{In Section \ref{__label__1d1140eeb1534e8b956d79bc5f640203}, we will simply take $\phi(\theta, \bs{Y}) = \|s(\bs{Y}) - s(\widehat{\bs{X}})\|$.} that describes the partial sample path $\bs{Y}$. We consider the (conditional) probability: 
\begin{equation}
 \rho \coloneqq \mathbb{P}\left[\|s(\bs{X}) - s(\widehat{\bs{X}}) \| < \varepsilon \mid \phi \right]. \label{__label__19b21599a08e4ad4879e128fa67ab1e4}
\end{equation} The conditional probability $\rho$ (which will need to be determined) indicates the chance that, based on the summary statistics of the partial sample, $\phi(\bs{Y})$, that the summary statistics of associated full (exact) sample path, $\bs{X}$, would be within an $\varepsilon$-distance of the test data, $\bs{\widehat{X}}$, and would consequently be \emph{accepted} by the ABC algorithm. We posit that the \emph{continuation probability}, $\alpha(\rho)$, should be such that:
\begin{enumerate}
 \item The function $\alpha(\rho)$ is increasing in $\rho$, so that sample paths more likely to meet the ABC acceptance criterion are preferentially continued.
 \item Further, $\alpha(\rho)$ is capable of taking a minimum value.
\end{enumerate} Writing $\alpha$ as $\alpha\left(\rho(\phi)\right)$, we can treat $\alpha: \phi \rightarrow [0, 1] $ as \begin{equation}
\alpha \left(\rho(\phi)\right) \coloneqq \min \Big[A \cdot \big[\rho(\phi)\big]^{B} + C, 1\Big], \hspace{6.9mm} (A, B, C \ge 0), \label{__label__8214bee2425743818c0a8773f17c425d}
\end{equation} where the values of $A$, $B$, and $C$ are chosen so that Equation \eqref{__label__c5426463d79b428197b24a63f7b08b9b} is maximal. The early-rejection ABC method is stated as Algorithm \ref{__label__213e85243fa84bf5b4adad47da3e84bb}.

 \begin{algorithm}[bth]
\caption{Early-rejection ABC\protect\vphantom{$A_A^A$}. The loop is repeated to return pairs of $(\theta, \omega$).}
\label{__label__213e85243fa84bf5b4adad47da3e84bb}
 \begin{algorithmic}[1]
  \Require data, $\widehat{\bs{X}}$, tolerance, $\varepsilon$, summary statistic, $s(\cdot)$, distance metric $\|\cdot\|$, and prior, $\pi(\cdot)$\protect\vphantom{$A_A^A$}
  \Require summary $\phi(\theta, \bs{Y})$ and continuation probability, $\alpha(\phi)$
\Loop
  \State sample $\theta \sim \pi(\cdot)$
  \State generate sample path $\bs{Y} \sim g(\cdot \mid \theta)$
  \State \textbf{with} probability $\alpha(\phi(\bs{Y}))$ \textbf{continue }
  \State\hspace{\algorithmicindent}generate sample path $\bs{X} \sim f(\cdot \mid \theta, \bs{Y})$
  \State calculate $\omega$ \begin{equation*}
  \omega = \begin{cases} \frac{1}{\alpha} \mathbb{I}\left\{\|s(\bs{X}) - s(\widehat{\bs{X}})\| < \varepsilon\right\} & \text { with probability } \alpha(\phi(\bs{Y})), \\ 
  
  0 & \text { otherwise. }\\\end{cases}
  \end{equation*}

\State \textbf{return} $(\theta, \omega)$
  \EndLoop
 \end{algorithmic}
\end{algorithm}

\subsection{Generating sampling paths with different resolutions} \label{__label__b27923208d3d4b08988f3e77aa5a3cb0}
In Section \ref{__label__2976342c969e4c8e8af17a17f45cf902} we explained that the stochastic process, defined by Equation \eqref{__label__84d2b77cf3bf4118b0e534292f4a7a43}, \begin{equation*} \bs{X}(T) = \bs{X}(0) + \sum_{j=1}^{M}  \mathcal{Y}_j \left(0,\int_0^T p_j(\bs{X}(t)) \wrt{t}\right) \cdot \nu_j,\end{equation*} can be simulated more efficiently if the tau-leap method is used. The tau-leap can be seen as making the substitution $\int_0^T p_j(\bs{X}(t)) \wrt{t} \rightarrow \sum_{k=0}^{K} p_j(\bs{X}(k\cdot\tau))\cdot \tau$. Thus, let the tau-leap process, $\bs{Y}$, be defined by Equation \eqref{__label__3576a62258ac4e9d8953e5bbea204ecf}, which is, \begin{equation*} \bs{Y}(K \cdot \tau) = \bs{Y}((K-1)\cdot \tau) + \sum_{j=1}^M \mathcal{Y}_j \left( \sum_{k=0}^{K-1} p_j(\bs{Y}(k\cdot\tau))\cdot \tau, \sum_{k=0}^{K} p_j(\bs{Y}(k\cdot\tau))\cdot \tau \right) \cdot \nu_j.\end{equation*} We let $\bs{Y}$ take the role of the partial sample path in Algorithm \ref{__label__213e85243fa84bf5b4adad47da3e84bb}. More generally, we let $\big\{\bs{Y_{\tau}} : \tau \in \{\tau_1 > \tau_2 > \dots \} \big\}$ be a family of partial sample paths. From Equation \eqref{__label__3576a62258ac4e9d8953e5bbea204ecf} it is clear that simulating a sample path $\bs{Y}_\tau$ is a matter of generating the Poisson processes, $\mathcal{Y}_j$ (for $j=1, \dots, M$). A Poisson process can be simulated in two ways: \begin{itemize}
\item The Poisson \emph{random number}, $\mathcal{P}(\Delta)$, indicates the number of arrivals in the Poisson process over an interval length $\Delta$ (but does not indicate the specific arrival times).
\item The exponential \emph{random number}, $\text{Exp}(1)$, indicates the waiting time between successive arrivals in the Poisson process.
\end{itemize} The tau-leap method generates sample paths quickly by using Poisson random numbers to `fire' multiple reactions at once; there is no time-saving (over the MNRM) if exponential random variables are simulated. 

As such, we start by generating $\bs{Y}_\tau$ with time-step $\tau \leftarrow \tau_1$. This means that, to advance from time $t = 0$ to time $t = K\cdot \tau$ in Equation \eqref{__label__3576a62258ac4e9d8953e5bbea204ecf}, we need to compute the number of arrivals in the time intervals $(0, p_j(\bs{Y}(0))\cdot\tau]$, $(p_j(\bs{Y}(0))\cdot\tau, p_j(\bs{Y}(\tau))\cdot\tau]$, $(p_j(\bs{Y}(\tau))\cdot\tau, p_j(\bs{Y}(2\tau))\cdot\tau]$, $\dots$, for the appropriate $j = 1, \dots, M$. Poisson random numbers can be used to generate these quantities.

The resolution of the sample path, $\bs{Y}_\tau$, can be increased if we change the time-step to $\tau' \leftarrow \tau_2$. In this case, we need to compute the number of arrivals in the time intervals $(0, p_j(\bs{Y}(0))\cdot\tau']$, $(p_j(\bs{Y}(0))\cdot\tau', p_j(\bs{Y}(\tau'))\cdot\tau']$, $(p_j(\bs{Y}(\tau'))\cdot\tau', p_j(\bs{Y}(2\tau'))\cdot\tau']$, $\dots$, for the appropriate $j = 1, \dots, M$. These do not correspond with the intervals previously generated, and so the existing Poisson process needs to be \emph{interpolated}. The following rules are used to interpolate such processes~\citep{__ref__a42fe24004244e1ca2f0aad5851f3653}:

\textbf{Rule 1}. If there are $K$ arrivals in a Poisson process over an interval $(\alpha, \gamma)$ then those arrivals are uniformly distributed over that interval. 

\textbf{Rule 2}. If there are $K$ arrivals in a Poisson process over an interval $(\alpha, \gamma)$ then, for $\beta$ between $\alpha$ and $\gamma$,  the number of arrivals over the interval $(\alpha, \beta)$ is binomially distributed as \begin{equation}
\mathcal{B} \left(K, \frac{\beta - \alpha}{\gamma - \alpha}\right). 
\end{equation}

\begin{figure}[!h]

\centering
\includegraphics[width=.75\linewidth]{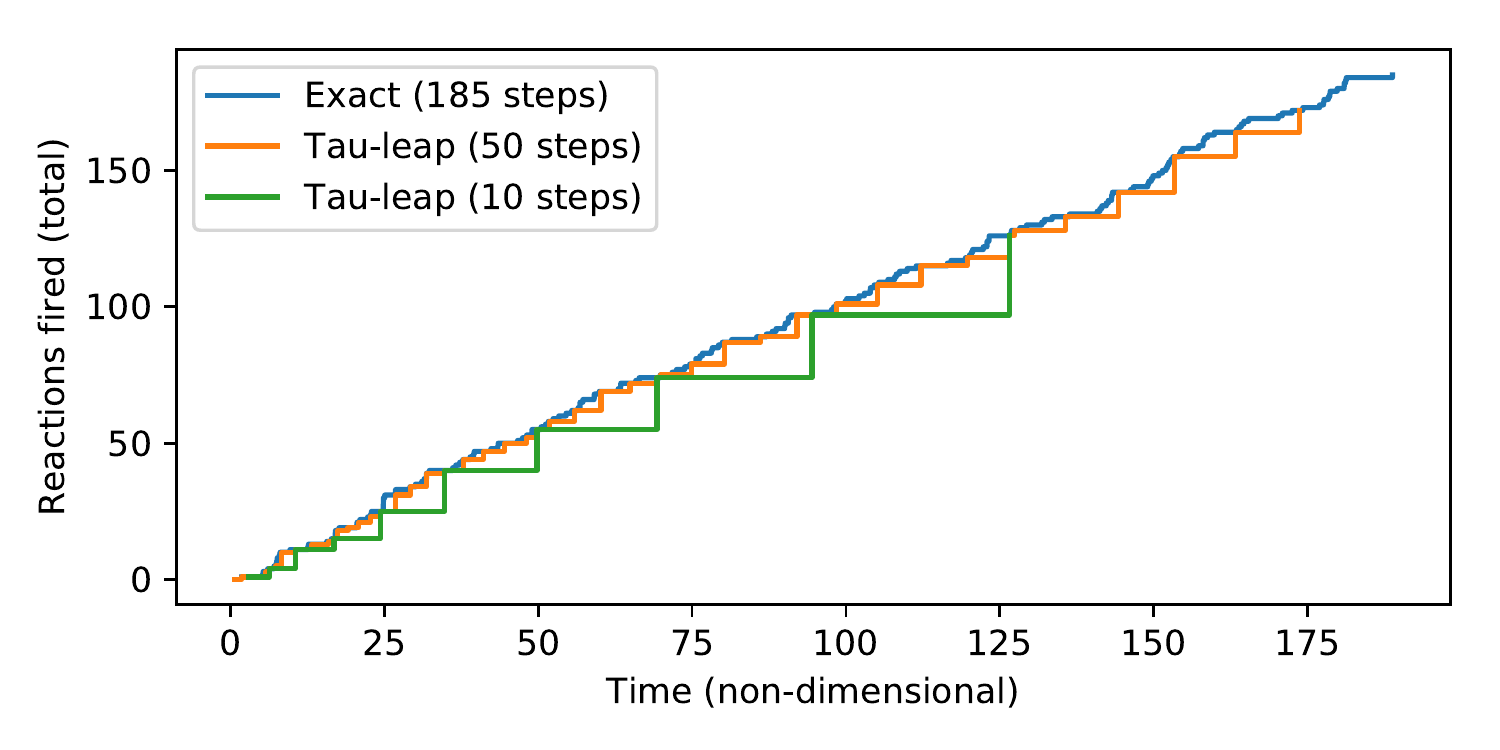} 
\caption{A unit-rate Poisson process is shown. The cumulative number of arrivals at different time-points is indicated. Initially, the Poisson process is not fully resolved as this is unnecessary. Where the number of arrivals needs to be calculated at different time-points, the process is interpolated per Rule 2.} \label{__label__4bdbd0b3aec7487fa1c8b5c534f77d97}

\end{figure}

Rule 1 allows a Poisson process to be prepared for the random input expected by the MNRM (Algorithm \ref{__label__17a38d1ae71849959c21f8e81c665c2b}), thereby making it possible to individually simulate each arrival of the Poisson process, and thereby generate an exact sample path, $\bs{X}$. Rule 2 is a corollary of Rule 1. Rule 2 ensures that a Poisson process can be efficiently interpolated by generating binomial random numbers, as it avoids the need to generate each arrival time with Rule 1. A return to our earlier case study illustrates the interpolation of a Poisson process:

\textbf{Case study.} A sample path of System \eqref{__label__5da48c70d96c4ee39993f234ef1294eb} was generated with multiple resolutions, and the results detailed in Figure \ref{__label__4ccf8701d0014669b51583e6f21de795}. In Figure \ref{__label__4bdbd0b3aec7487fa1c8b5c534f77d97} we show how a single Poisson process is repeatedly interpolated to generate the sample path (described in Figure \ref{__label__4ccf8701d0014669b51583e6f21de795}) with different resolutions. 

\subsection{The ML-ABC algorithm} \label{__label__786cad91bb664bc78e7876eda9a9a892}
We are in a position to set out the ML-ABC method. First, we settle on a choice of approximate sample resolutions, and therefore the set $\big\{\bs{Y}_{1}, \dots, \bs{Y}_{L}: \text{ where } \bs{Y}_\ell \text{ has time-step } \tau_\ell\big\}$. Then, \begin{itemize}
\item Sample $\theta \sim \pi(\cdot)$ and label the Poisson processes as $\mathcal{Y}_j$ for $j = 1, \dots, M$.
\item Generate $\bs{Y}_{1}$ with time-step $\tau_1$ .
\end{itemize} With probability $\alpha_1(\phi_1(\bs{Y}_{1}))$ proceed to (using the aforementioned Poisson processes): \begin{itemize}
\item Generate $\bs{Y}_{2}$ with time-step $\tau_2$.
\end{itemize}With conditional probability $\alpha_2(\phi_2(\bs{Y}_{1}, \bs{Y}_{2}))$ proceed to (using the  Poisson processes): \begin{itemize}
\item Generate $\bs{Y}_{3}$ with time-step $\tau_3$.
\end{itemize} Proceed recursively, until, with conditional probability $\alpha_L(\phi_L(\bs{Y}_{1}, \dots, \bs{Y}_{L}))$ we proceed to (using the Poisson processes): \begin{itemize}
\item Generate $\bs{X}$ with the MNRM. $\bs{X}$ is the \emph{exact} or complete sample path.
\end{itemize} A pair $(\theta, \omega)$ is then returned; the weight, $\omega$, is given by \begin{equation}
\omega = \begin{cases}
          \Big[\prod \alpha_i \Big]^{-1} \mathbb{I}\left\{\|s(\bs{X}) - s(\widehat{\bs{X}})\| < \varepsilon\right\}  & \text{ if } \bs{X} \text { is generated,} \\
          0 & \text{ otherwise.} \\
          \end{cases}
         \end{equation}
         
In practice, we suggest the following procedure for choosing $\alpha_\ell$. A small number of survey simulations are generated, with $\alpha_1 = \dots = \alpha_L = 1$. Then, we take $\phi_\ell$ to be the current `error', given by: \begin{equation} \phi_\ell(\bs{Y}_{1}, \dots, \bs{Y}_\ell) = \|s(\bs{Y}_\ell) - s(\widehat{\bs{X}})\|. \label{__label__6bbb0ccde03e427e93ad36e546da6522} \end{equation} For clarity, we label the quantity \eqref{__label__6bbb0ccde03e427e93ad36e546da6522} as $\text{err}(\bs{Y}_\ell)$. In Section \ref{__label__0ae2c4ede8e94b68a908d2e812d70329}, Equation \eqref{__label__19b21599a08e4ad4879e128fa67ab1e4} sets out the probability, should the current path $\bs{Y}_\ell$ be upgraded to its associated exact sample path, $\bs{X}$, that this exact sample path would be within an $\varepsilon$-distance of the test data. Specifically, $$\rho_\ell = \mathbb{P}\left[\|s(\bs{X}) - s(\widehat{\bs{X}}) \| < \varepsilon \mid \|s(\bs{Y}_\ell) - s(\widehat{\bs{X}}) \|\right].$$ The conditional probabilities, $\rho_\ell (\phi_\ell) $ can be estimated in many ways -- for example, it could be sufficient to fit a decision function $\rho_\ell(\xi) = \exp\big[\sum_{i} \lambda_i(\xi - \xi_0)^i\big]$, together with a suitable regularisation, to the survey simulation data.

Once the conditional probability functions, $\rho_\ell(\phi_\ell)$, have been estimated, the continuation probabilities, $\alpha_\ell(\phi_\ell)$, can be constructed in accordance with Equation \eqref{__label__8214bee2425743818c0a8773f17c425d}. In particular, we use the survey simulations, together with a a brute force search to choose parameters, $A_\ell$, $B_\ell$ and $C_\ell$, so that $$\alpha_\ell(\phi_\ell) \coloneqq \min \Big[A_\ell \cdot \big[\rho_\ell(\phi_\ell)\big]^{B_\ell} + C_\ell, 1\Big],$$ with Equation \eqref{__label__c5426463d79b428197b24a63f7b08b9b} maximized.

\section{Case studies} \label{__label__1d1140eeb1534e8b956d79bc5f640203}
\begin{figure}[ht]

\centering
\includegraphics[width=\linewidth]{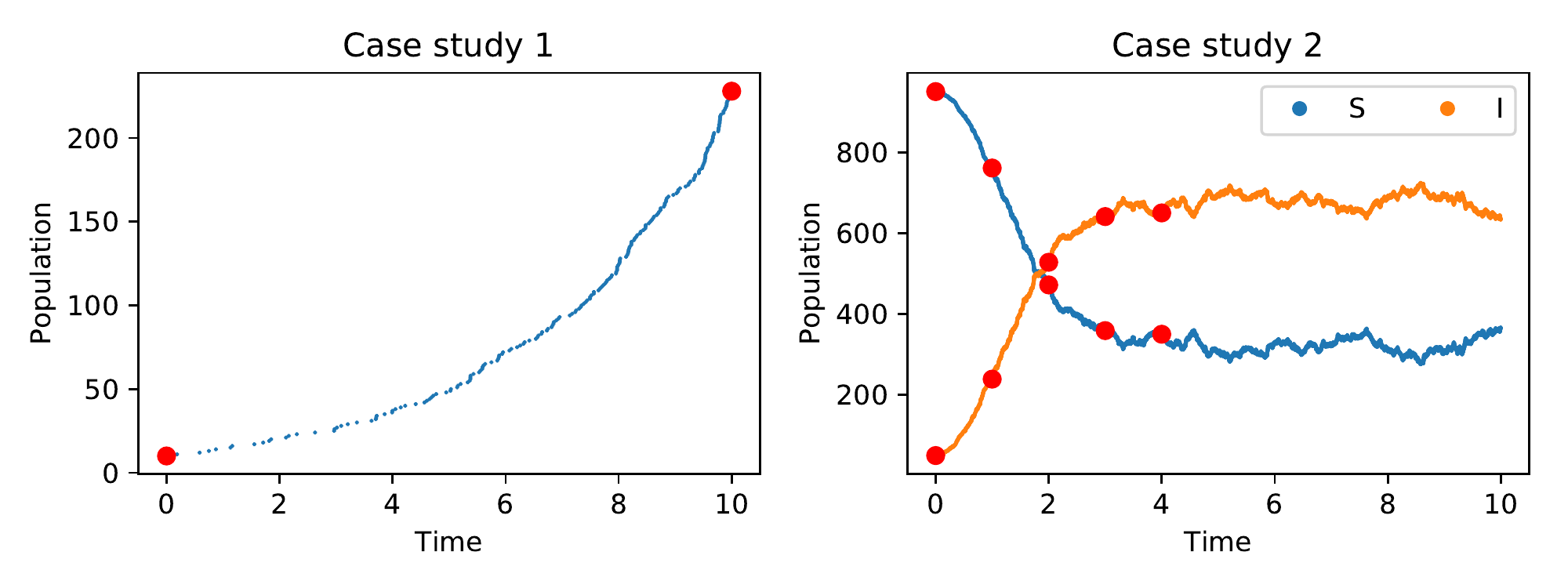} 
\caption{On the left, an \emph{in-silico} data path of a birth process is shown. The data path on the right is of an S-I-S model. The data were randomly generated, with the random seed set to the author's birth-date. The proposed population observations are shown in red.} 

\label{__label__96677bd565644bbfb1888b5253545ba6}

\end{figure}

In this section, we present two case studies; a discussion of our results then follows in Section \ref{__label__a48aec03dc24496dba274350bb9a76b0}. The first case study is concerned with a birth process (Section \ref{__label__7a1ce86a413d41e0a847bcf46a2614c0}), and the second case study models the spread of a disease through an S-I-S compartment model (Section \ref{__label__1b5a5bf3657f497b9120e9e800ee1ffd}). Figure \ref{__label__96677bd565644bbfb1888b5253545ba6} presents the \emph{in-silico} data that we have generated to test the ML-ABC algorithm on. For each of Case studies 1 and 2, a single time-series is shown; all the sample paths were generated on an Intel Core i5 CPU rated at 2.5 GHz.

\subsection{Case study 1: a birth process} \label{__label__7a1ce86a413d41e0a847bcf46a2614c0}
The first case study is concerned with a birth process. As in Section \ref{__label__6b90150fcf9547afa4c1f2a5cac7059f}, this process comprises a single species, $X$, together with a reaction channel $R_1$: \begin{equation*} R_1: X \xrightarrow{\theta} X + X, \end{equation*} with $X(0) = 10$. We seek to infer the parameter $\theta$. The data, $\bs{\widehat{X}}$, are shown in Figure \ref{__label__96677bd565644bbfb1888b5253545ba6} (note that $X(0)$ is chosen in accordance with Figure \ref{__label__96677bd565644bbfb1888b5253545ba6}). In particular, for our purposes it is sufficient to let the summary statistic, $s$, be given by the final population, $s(\bs{X}) = \bs{X}(10)$. 

\begin{figure}[ht]

\centering
\includegraphics[width=.75\linewidth]{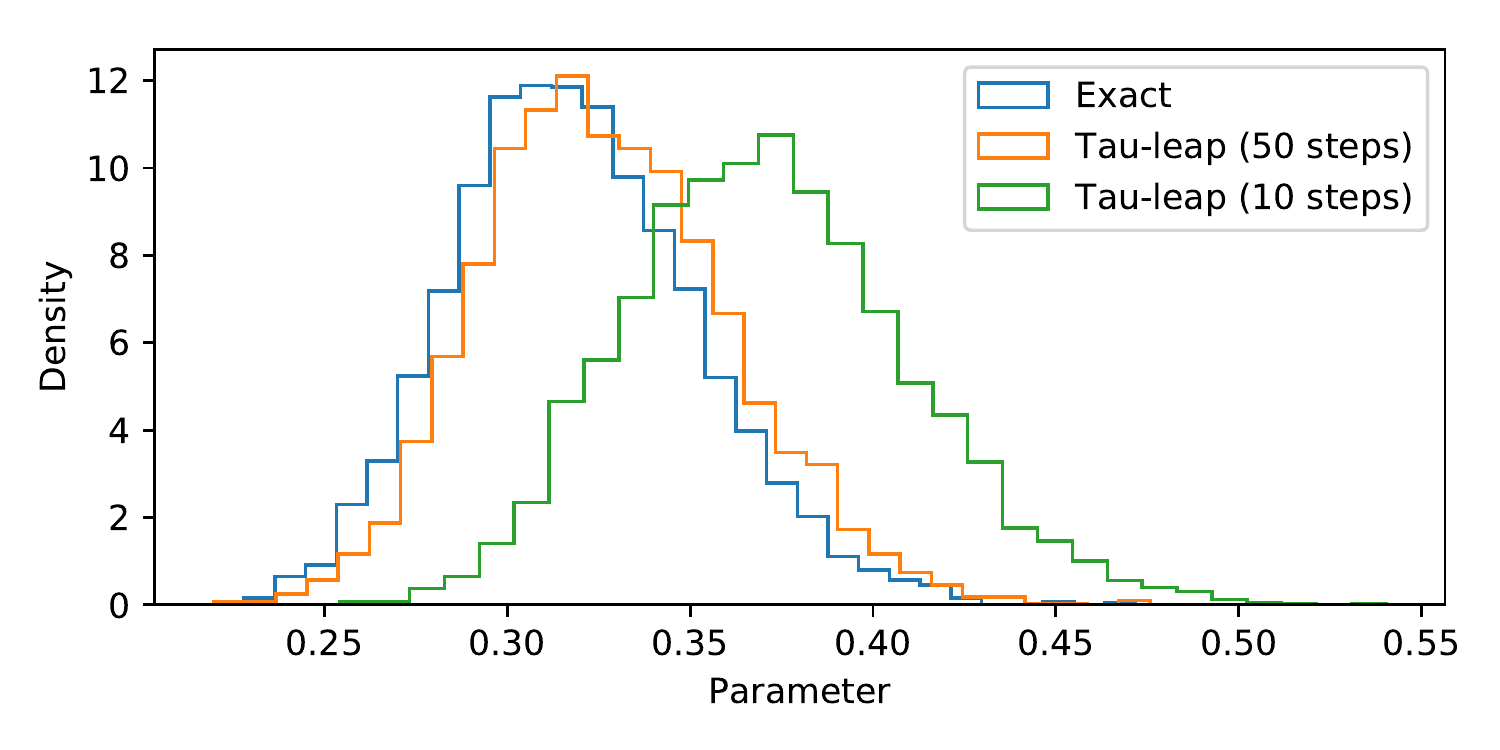} 
\caption{Empirical posterior distributions for the parameter $\theta$ of System \eqref{__label__5da48c70d96c4ee39993f234ef1294eb} are shown. Different simulation methods are used; a parameter of $\varepsilon < 35$ is used to decide whether to accept a sample. Approximately $3.1\%$ of MNRM sample paths are accepted; where the tau-leap method is used, $4.2\%$ of sample paths are accepted where $\tau = 1.0$, and where $\tau = 0.2$, the rate is $3.3\%$.}

\label{__label__4b71f0c2d9734e098338bb1d63221ae9}

\end{figure}

We use a uniform prior distribution, taking $\pi(\cdot) \sim \mathcal{U}[0.01, 1.00]$. We then generate $N = 10^5$ sample paths using the MNRM (see Algorithm \ref{__label__17a38d1ae71849959c21f8e81c665c2b}). The error associated with each sample path is given by \begin{equation}
\text{err}(\bs{X}) \coloneqq \|s(\bs{X}) - s(\bs{\widehat{X}}) \| = |\bs{X}(10) - \bs{\widehat{X}}(10)|. \label{__label__f6043cfa2b8a46ca9fa00aa50056aa2a}
\end{equation} Taking a tolerance of $\varepsilon = 35$, an empirical posterior distribution is plotted, and the results detailed in Figure \ref{__label__4b71f0c2d9734e098338bb1d63221ae9}. The maximum of this empirical posterior distribution is ${\theta} = 0.304$, which compares very favourably with the choice of $\widehat{\theta} = 0.3$ used to generate the \emph{in-silico} data. In addition, Figure \ref{__label__4b71f0c2d9734e098338bb1d63221ae9} indicates the empirical posterior distributions associated with the use of tau-leap sample paths, with $\tau$ taken as $\tau = 1.0$ and $\tau = 0.2$. This figure is consistent with the tau-leap sample paths introducing an additional bias.

The ML-ABC method is now implemented. To illustrate the algorithm, we take $\tau_1 = 1.0$ and $\tau_2 = 0.2$. As outlined in Section \ref{__label__659100d1f2c14b66a0f4493f0729ea9b}, we sample $\theta \sim \pi(\cdot)$, and we then generate a sample path with time-step $\tau_1 = 1.0$, $\bs{Y}_{1}$. If $\bs{Y}_{1}$ is such that it is likely to contribute to the empirical posterior distribution, then we re-generate $\bs{Y}_{1}$, but with a smaller time-step of $\tau_2 = 0.2$. As the same Poisson processes are used, $\bs{Y}_{2}$ is a higher-resolution version of $\bs{Y}_{1}$. Based on $\bs{Y}_{2}$, we decide whether the exact (or complete) sample path, $\bs{X}$ should be generated. At each stage, we re-weight our sample to avoid introducing an additional bias. 

\begin{figure}[ht]
\centering
\includegraphics[width=\linewidth]{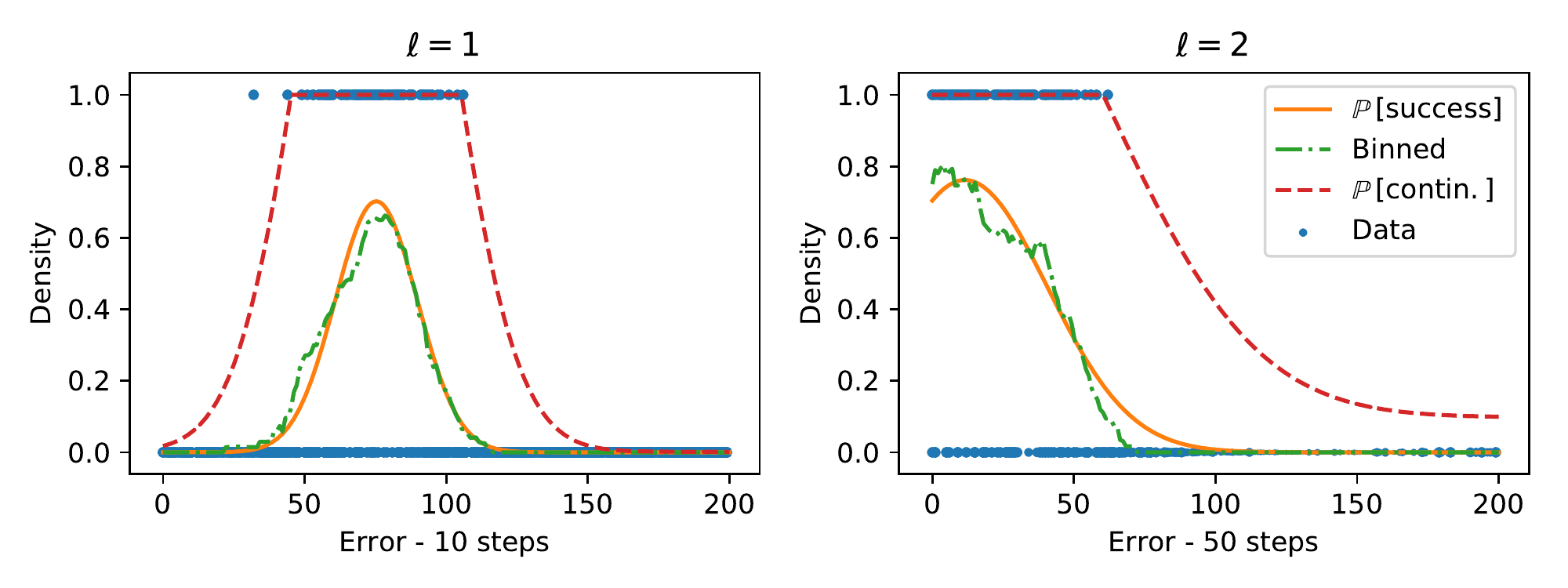} 
\caption{On the left, we plot an indicator function $\mathbb{I}\{|s(\bs{X}) - s(\bs{\widehat{X}}) |< \varepsilon\}$ against the error $\text{err}(\bs{Y}_{1})$ in blue. A moving average is shown in green, and a Gaussian-form curve is fitted in orange. The curve indicates the probability, given $\text{err}(\bs{Y}_{1})$, that if the complete sample $\bs{X}$ is generated from $\bs{Y}_{1}$, that the complete sample will be within an $\varepsilon$-distance of the test data: this is the function $\rho$ specified by Equation \eqref{__label__19b21599a08e4ad4879e128fa67ab1e4}. The red line shows the continuation probability, $\alpha$, as specified by Equation \eqref{__label__36d5d9bd50084db38ff3084419d9c8ff}. On the right, an equivalent graph for $\text{err}(\bs{Y}_{2})$ is shown.} 
\label{__label__c624f12d4a1942ba8a02516a670264de}

\end{figure}

In order to decide which sample paths to refine, we first generate a number of survey simulations to calibrate the model. Thus, we generate $N'$ triples $\{\bs{Y}_{1}, \bs{Y}_{2}, \bs{X}\}$. The value of $N'$ can be specified directly, but we will continue generating sample paths until at least $100$ of the $\bs{X}$ sample paths are \emph{accepted}\footnote{In this case, $N'=3\,056$.}. As explained in Section \ref{__label__0ae2c4ede8e94b68a908d2e812d70329}, any suitable machine-learning classifier can then be used to probabilistically associate the error, $\text{err}(\bs{Y}_{1})$ (see Equation \eqref{__label__f6043cfa2b8a46ca9fa00aa50056aa2a}) with the final outcome, $\mathbb{I}\{\text{err}(\bs{X}) < \varepsilon\}$, that is the probabilistic relationship (see Equation \eqref{__label__19b21599a08e4ad4879e128fa67ab1e4}): \begin{equation}
\text{err}(\bs{Y}_{1}) = |s(\bs{Y}_{1}) - s(\bs{\widehat{X}}) | \,\,\,\text{ to the class }\,\,\, \mathbb{I}\{|s(\bs{X}) - s(\bs{\widehat{X}}) |< \varepsilon\}.
\end{equation} To this end, we use a simple, Gaussian-form probability function, which is fitted to the test data. The function, $\rho_1$, is indicated on the left-side of Figure \ref{__label__c624f12d4a1942ba8a02516a670264de}. Following from Equation \eqref{__label__8214bee2425743818c0a8773f17c425d}, it is sufficient for our continuation probability, $\alpha_1$, to be taken as \begin{equation}
\alpha_1 \coloneqq \min \Big[A_1 \cdot \big[\rho_1(\text{err}(\bs{Y}_{1}))\big]^{B_1} + C_1, 1\Big], \label{__label__36d5d9bd50084db38ff3084419d9c8ff}
\end{equation} where $A_1$, $B_2$ and $C_3$ are to be determined later.

Having specified the form of $\alpha_1$, our attention turns to $\alpha_2$. Whilst Section \ref{__label__659100d1f2c14b66a0f4493f0729ea9b} indicates more general formulations are possible, we seek to probabilistically relate \begin{equation}
\text{err}(\bs{Y}_{2}) = |s(\bs{Y}_{2}) - s(\bs{\widehat{X}}) | \,\,\,\text{ to the class }\,\,\, \mathbb{I}\{|s(\bs{X}) - s(\bs{\widehat{X}}) |< \varepsilon\}.
\end{equation} We use another Gaussian-form probability function, which is fitted to the test data. This decision function is labelled as $\rho_2$. The data are to be weighted in proportion to the previous continuation probability, $\alpha_1$. This means that samples that are unlikely to continue to the second stage should have a relatively small effect on $\rho_2$. The right-side of Figure \ref{__label__c624f12d4a1942ba8a02516a670264de} indicates the second decision function. As before, we now take \begin{equation}
\alpha_2 \coloneqq \min \Big[A_2 \cdot \big[\rho_2(\text{err}(\bs{Y}_{2}))\big]^{B_2} + C_2, 1\Big]. \label{__label__31b4237e8953428faff9a3b1b35b55e3}
\end{equation} The values of $A_\ell$, $B_\ell$ and $C_\ell$, for $\ell \in \{1, 2\}$, are optimised over the set of survey simulations. An iterative procedure is followed\footnote{Note that, if $A_1$, $B_1$ and $C_1$ are yet to be chosen, then $\alpha_1$ is unspecified, and the data weightings required to determine $\rho_2$ are unavailable.}: (1) set $A_\ell$, $B_\ell$ and $C_\ell$ ($\forall \ell$) to initial values; (2) calculate $\rho_\ell$ and $\alpha_\ell$ ($\forall \ell$) with the current values of $A_\ell$, $B_\ell$ and $C_\ell$, using appropriate data weightings; (3) optimise $A_\ell$, $B_\ell$ and $C_\ell$ ($\forall \ell$); then, repeat (3) and (4) as required. 

It is now possible to run the full algorithm, with $N = 75\,000$ initial samples of $\bs{Y}_{1}$. Of the $N = 75\,000$ samples, $12\%$ are selected to be refined into $\bs{Y}_{2}$. Of those $9\,239$ samples, only $61\%$ of samples are further refined to the exact sample path, $\bs{X}$. Of the $5\,644$ active samples, $2329$ or $41\%$ are accepted (as $\text{err}(\bs{X}) < \varepsilon$), and contribute to the empirical posterior distribution. The ESS (see Equation \eqref{__label__045cbdb1be72461b96f5c0175bb68360}) takes into account sample weights, meaning that the ESS is $2\,314$. The effective samples per CPU-second is $3\,297.5$ units.

If the regular, ABC-rejection method (see Algorithm \ref{__label__678f924617094025815fa9832fc0f45c}) is used, with $N= 75\,000$ initial samples and $\varepsilon = 35$, then $3\,107$ samples are accepted for inclusion in the posterior distribution. This is a rate of $16.9$ effective samples per CPU-second. The result is that the ML-ABC method is approximately $195$ times more efficient than rejection ABC.

\subsection{Case study 2: an S-I-S model} \label{__label__1b5a5bf3657f497b9120e9e800ee1ffd}
In this second case study, we consider an S-I-S infection model~\citep{__ref__dde7e7d3ceb3449e81a5e69b45bc72e5}. The model comprises two species, $S$ (the `susceptible' individuals) and $I$ (the `infected' individuals). Initially, $S = 950$ and $I = 50$ (these quantities can be deduced from Figure \ref{__label__96677bd565644bbfb1888b5253545ba6}). There are two reaction channels: \begin{equation}
R_1: S + I \xrightarrow{\theta_1} 2 \cdot I \quad\quad \quad  R_2: I \xrightarrow{\theta_2} S. \label{__label__8e9dc5f0670f48148a591b51e01e67e8}
\end{equation} We seek the values of parameters $\theta_1$ and $\theta_2$. The data, $\bs{\widehat{X}}$, are shown in Figure \ref{__label__96677bd565644bbfb1888b5253545ba6}. In particular, for our purposes it is sufficient to let the summary statistic, $s$, be given by the vector, $s(\bs{X}) = [S(1), \dots, S(4), I(1), \dots, I(4)]$. 

\begin{figure}[bht]
\centering
\includegraphics[width=\linewidth]{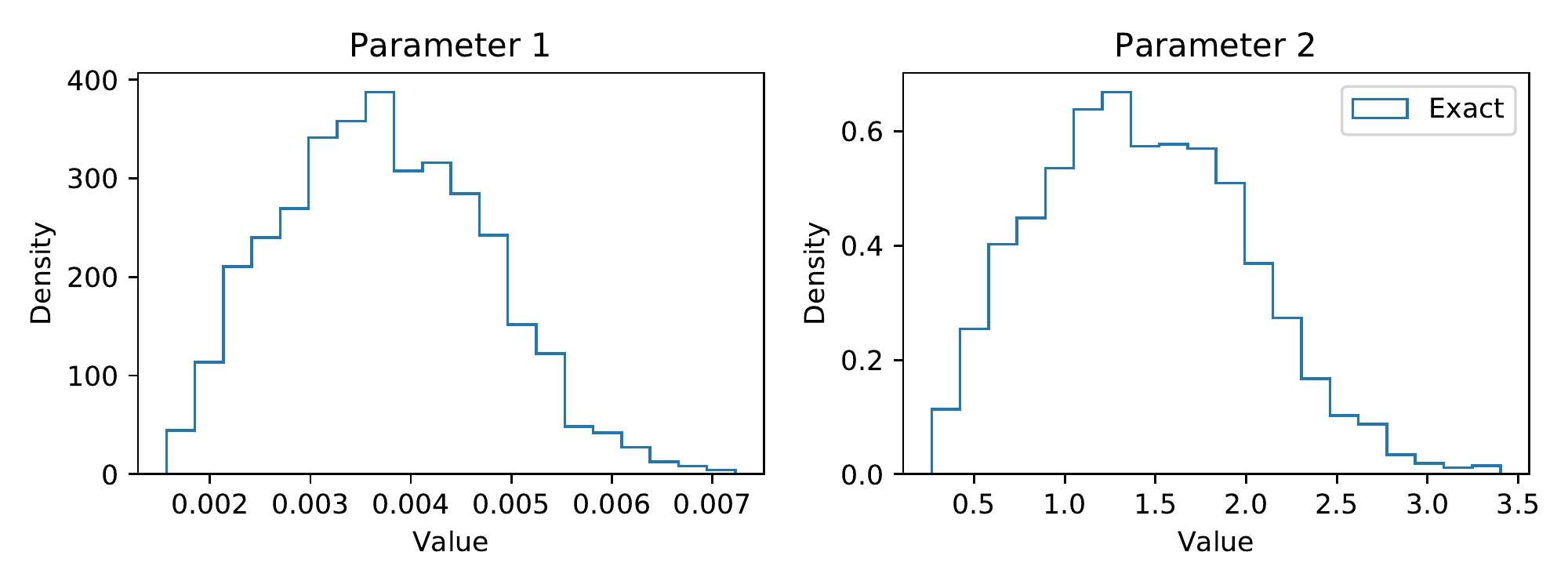} 
\caption{Empirical posterior distributions for the parameters $\theta_1$ and $\theta_2$ of System \eqref{__label__8e9dc5f0670f48148a591b51e01e67e8} are shown. A parameter of $\varepsilon < 250$ is used to decide whether to accept a sample. Approximately $1.7\%$ of the sample paths have been accepted.} 
\label{__label__6f5679239c80441f8be9750246b3f2dd}

 \end{figure}

We use a uniform prior distribution, taking $\pi(\cdot) \sim \mathcal{U}[0.0001, 0.02] \times \mathcal{U}[0.0001, 5.0]$. We generate $N = 10^5$ sample paths using the MNRM (see Algorithm \ref{__label__17a38d1ae71849959c21f8e81c665c2b}). The error associated with each sample path is given by the $\ell^2$-distance \begin{equation}
\text{err}(\bs{X}) \coloneqq \|s(\bs{X}) - s(\bs{\widehat{X}}) \|_2. \label{__label__dcbb6531ddd14e60a0d3f1e27b2e08da}
\end{equation} Taking a tolerance of $\varepsilon = 250$, an empirical posterior distribution is plotted, and the results detailed in Figure \ref{__label__6f5679239c80441f8be9750246b3f2dd}. The maximums of the point-wise empirical posterior distributions are $(\theta_1, \theta_2) = (0.0035, 1.207)$, which compares with the choice of $(\widehat{\theta_1}, \widehat{\theta_2}) = (0.003, 1.000)$ used to generate the \emph{in-silico} data. 

The ML-ABC method is now implemented. Given the larger domain spanned by the prior, $\pi(\cdot)$, we will use an adaptive tau-leap method to generate $\bs{Y}_{1}$. This means that, instead of specifying the time-step $\tau$ directly, we specify a parameter, $\xi$, that parameterises a tau-selection algorithm. In this case, the precise value of $\tau$ is chosen according to $\xi$ and the state vector (see Equation \eqref{__label__9a73ab33264847b5bd60742aaf26dc86}). We use the popular tau-selection method of \citet{__ref__6c34939f18e54a3a8af80284359d4e56}. We use only one approximation level, so to generate $\bs{Y}_{1}$ we fix $\xi = 0.2$. If $\bs{Y}_{1}$ is such that it is likely to contribute to the empirical posterior distribution, then we generate the complete sample path, $\bs{X}$. The Poisson processes are interpolated according to Rule 1, and Algorithm \ref{__label__17a38d1ae71849959c21f8e81c665c2b} is then run. 

\begin{figure}[ht]
\centering
\includegraphics[width=.75\linewidth]{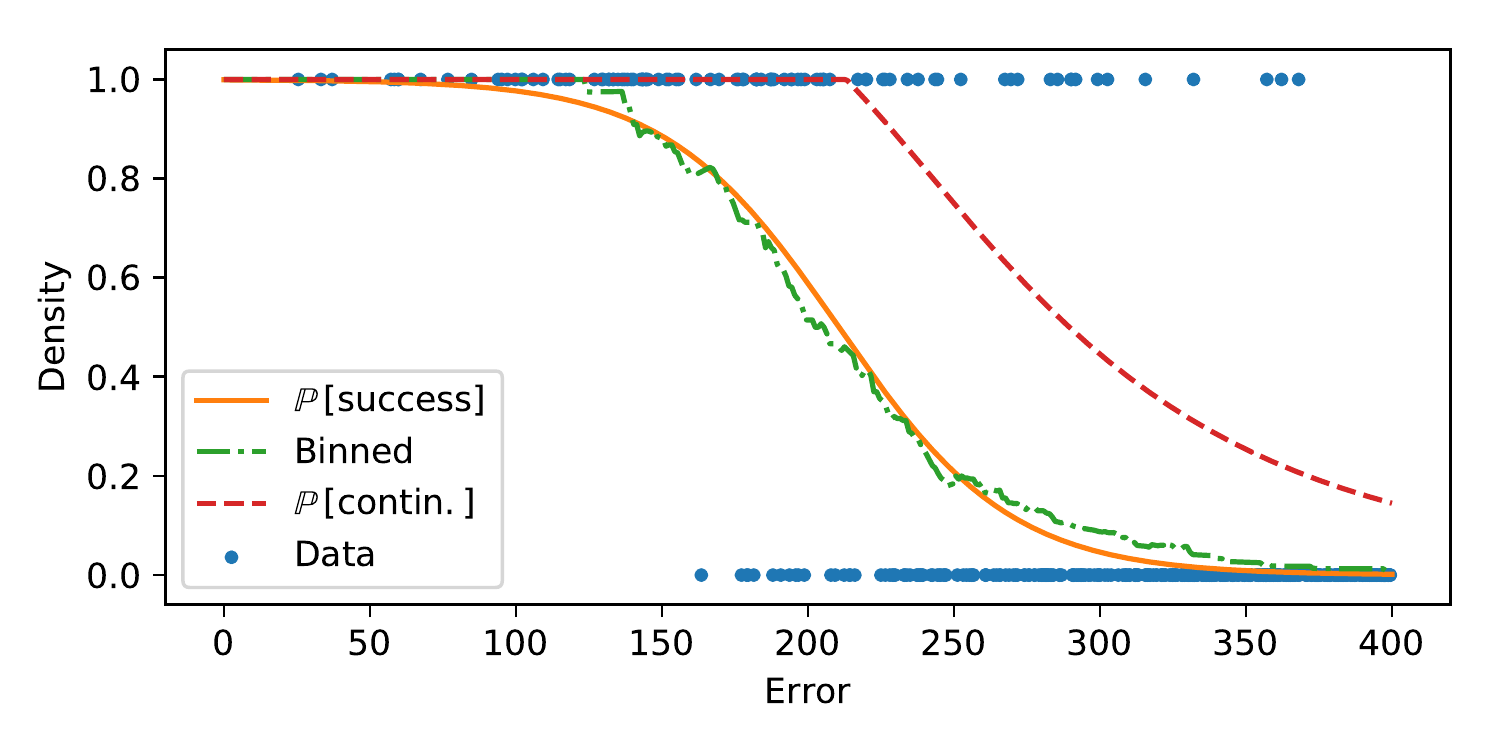} 
\caption{We plot an indicator function $\mathbb{I}\{|s(\bs{X}) - s(\bs{\widehat{X}}) |< \varepsilon\}$ against the error $\text{err}(\bs{Y}_{1})$ in blue. A moving average is shown in green, and a logistic regression curve is fitted for $\rho$ in shown orange. The red line shows the continuation probability, $\alpha$, as specified by Equation \eqref{__label__b7f7b36a01df4b4fa14a10e54ac760fb}.} 
\label{__label__c08d244625de49acb042f5382b9f8f40}

 \end{figure}

As before, we must first generate a number of survey simulations to calibrate the model. A total of $N' = 6\,404$ survey simulations are required to assimilate $100$ sample paths, $\bs{X}$, that are \emph{accepted} by the algorithm, and, therefore, we have generated $N'$ pairs $\{\bs{Y}_{1}, \bs{X}\}$. As before, any suitable machine-learning classifier can then be used to associate the error, $\text{err}(\bs{Y}_{1})$ (see Equation \eqref{__label__f6043cfa2b8a46ca9fa00aa50056aa2a}) with the final outcome, $\mathbb{I}\{\text{err}(\bs{X}) < \varepsilon\}$. In this case, we use logistic regression to estimate the conditional probability, $\rho_1(\phi)$ (where $\phi = \text{err}(\bs{Y}_1)$). Our logistic regression is shown in Figure \ref{__label__c08d244625de49acb042f5382b9f8f40}.

 Following from Equation \eqref{__label__8214bee2425743818c0a8773f17c425d}, it is sufficient for our continuation probability, $\alpha$, to be taken as \begin{equation}
\alpha \coloneqq \min \Big[A_1 \cdot \big[\rho_1(\text{err}(\bs{Y}_{1}))\big]^{B_1} + C_1, 1\Big], \label{__label__b7f7b36a01df4b4fa14a10e54ac760fb}
\end{equation} where $\rho_1$ is the logistic regression function described above, and $A_1$, $B_1$ and $C_1$ are optimised on the survey simulations. 

It is now possible to run the full algorithm, with $N = 75\,000$ initial samples of $\bs{Y}_{1}$. Of the $N = 75\,000$ samples, $5\%$ are selected to be refined into exact samples, $\bs{X}$. Of the $3\,804$ samples, $1\,091$ or $29\%$ are accepted, and contribute to the empirical posterior distribution. The ESS (see Equation \eqref{__label__045cbdb1be72461b96f5c0175bb68360}) takes into account sample weights; the ESS is recorded as $955.0$. The effective samples per CPU-second is $135.3$ units.

If the regular, ABC-rejection method (see Algorithm \ref{__label__678f924617094025815fa9832fc0f45c}) is used, with $N= 75\,000$ initial samples and $\varepsilon = 250$, then $1\,677$ samples are accepted for inclusion in the posterior distribution. This is a rate of $17.0$ effective samples per CPU-second. The result is that the ML-ABC method is approximately $7.9$ times more efficient than rejection ABC.

\subsection{Discussion} \label{__label__a48aec03dc24496dba274350bb9a76b0}
Stochastic models of complicated real-world phenomena often rely on carefully chosen parameters. As such, the design and development of efficient inference algorithms play a pivotal role in underpinning the work of the modelling community. In this work, we presented the ML-ABC method, and demonstrated its efficacy at inferring model parameters. In particular, we demonstrated a multi-resolution inference method, where a machine-learning-led approach can select sample paths for further refinement, without introducing an additional bias. We conclude with an \emph{obiter dictum} regarding decision functions.

\begin{figure}[ht]
\centering
\includegraphics[width=\linewidth]{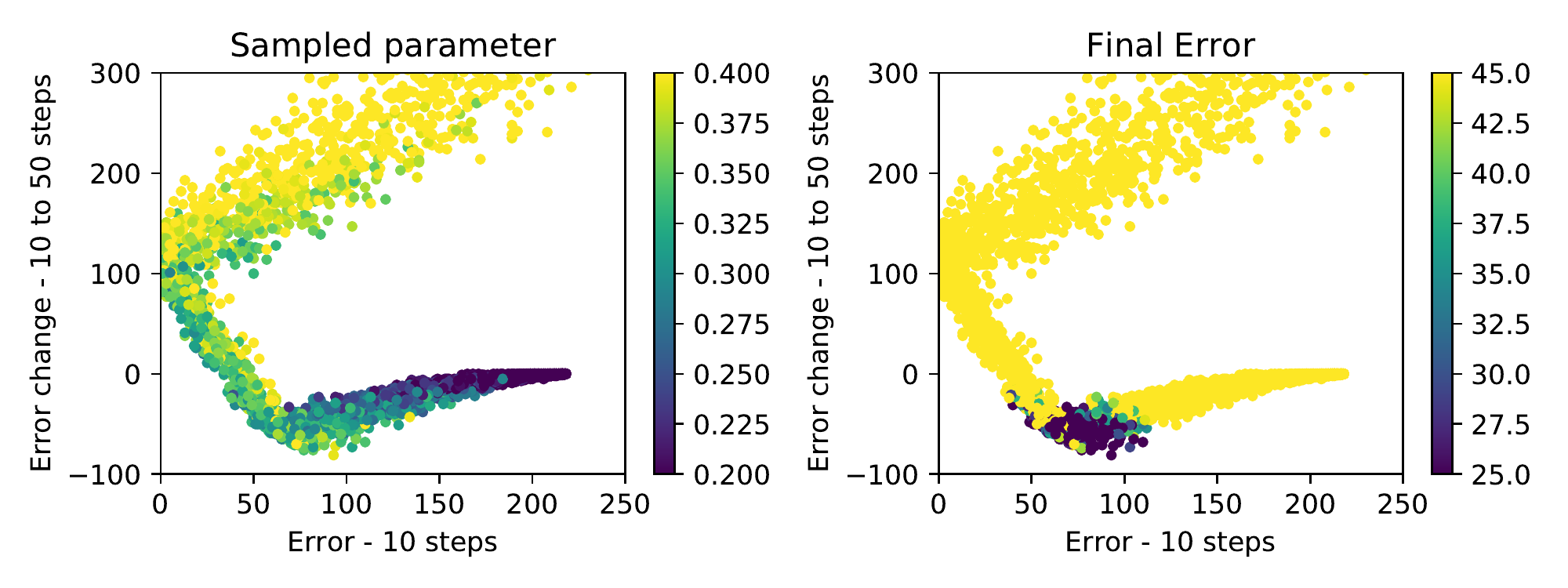} 
\caption{On the $x$-axis, we indicate $\text{err}(\bs{Y}_{1})$, and on the $y$-axis, we show $\big[\text{err}(\bs{Y}_{2}) - \text{err}(\bs{Y}_{1})\big]$. The left-side graph shows a data point for each triple, $\{\bs{Y}_{1}, \bs{Y}_{2}, \bs{X}\}$, with the colour of the data-point corresponding to the parameter, $\theta$, sampled from the prior. The right-side shows the same data points, but with the colour corresponding to the final error, $\|s(\bs{X}) - s(\bs{\widehat{X}}) \|$. Approximately $43\%$ of data points are within the axes limits. The colour-bars have been truncated.} 
\label{__label__e1619c37d10b4de695d946ab74fd2c6c}

\end{figure}

\textbf{Decision functions.} In Section \ref{__label__7a1ce86a413d41e0a847bcf46a2614c0}, we presented a case study of a birth process. In Figure \ref{__label__e1619c37d10b4de695d946ab74fd2c6c}, we consider, $\text{err}(\bs{Y}_{1})$ and $\big[\text{err}(\bs{Y}_{2}) - \text{err}(\bs{Y}_{1})\big]$. The first quantity indicates the distance between the summary statistics of $\bs{Y}_{1}$ and the data, $\bs{\widehat{X}}$, whilst the second quantity indicates the \emph{change} in the error, as $\tau$ is reduced from $\tau = 1.0$ to $\tau = 0.2$. We show how the aforementioned quantities are related to (a) the parameter that has been drawn from the prior, $\theta$; and (b) the final error, that is $\|s(\bs{X}) - s(\bs{\widehat{X}}) \|$. Under certain circumstances, it might be favourable to generate \emph{both} $\bs{Y}_{1}$ and $\bs{Y}_{2}$, and then, make a single decision as to whether one should continue to generate the complete sample, $\bs{X}$. 

In our view, the future of Approximate Bayesian Computation is inextricably tied to machine learning approaches that probabilistically select sample paths and parameters for further investigation. 

\newpage

\bibliographystyle{v2}
\bibliography{b1_clean.bib}

\end{document}